\documentclass{llncs}[]
\usepackage{amssymb}
\usepackage{graphicx}
\usepackage{url}
\usepackage{multirow}
\pdfoutput=1

\begin{document}

\title{Towards platform-independent \\
specification and verification \\ of the standard trigonometry functions\thanks{This research is supported by Russian Basic Research Foundation grant
no. 17-01-00789 \emph{Platform-independent approach to formal specification and verification of standard mathematical functions}.}}

\titlerunning{Towards verification of the square root}

\author{Nikolay V. Shilov\inst{1}, Boris L. Faifel\inst{2}, Svetlana O. Shilova\inst{3}, Aleksey V. Promsky\inst{4}}

\authorrunning{N. Shilov et. al.}

\institute{Innopolis University, Innopolis, Russia, \email{shiloviis@mail.ru},\\
\and Yu. Gagarin State Technical University of Saratov, Russia, \email{catstail@yandex.ru}, \\
\and retired, Novosibirsk, Russia \email{shilov61@inbox.ru},
\and A.P. Ershov Institute of Informatics Systems, Novosibirsk, Russia \email{promsky@iis.nsk.su}}

\maketitle

\begin{abstract}
Research project ``Platform-independent approach to formal specification and verification of standard mathematical functions''
is aimed onto a development of an incremental combined approach to the specification and verification of the standard mathematical functions
like \texttt{sqrt}, \texttt{cos}, \texttt{sin}, etc.
Platform-independence means that we attempt to design a relatively simple axiomatization of the computer arithmetic in terms of real, rational, and integer arithmetic
(i.e. the fields $\mathbb{R}$ and $\mathbb{Q}$ of real and rational numbers, the ring $\mathbb{Z}$ of integers)
but don’t specify neither base of the computer arithmetic, nor a format of numbers’ representation.
Incrementality means that we start with the most straightforward specification of the simplest easy to verify algorithm in real numbers and
finish with a realistic specification and a verification of an algorithm in computer arithmetic.
We call our approach combined because we start with a manual (pen-and-paper) verification of some selected algorithm in real numbers,
then use these algorithm and verification as a draft and proof-outlines for the algorithm in computer arithmetic and its manual verification,
and finish with a computer-aided validation of our manual proofs with some proof-assistant system
(to avoid appeals to ``obviousness'' that are very common in human-carried proofs).
In the paper we present first steps towards a platform-independent incremental combined approach
to specification and verification of the standard  functions \texttt{cos} and \texttt{sin}
that implement mathematical trigonometric functions $\cos$ and $\sin$.

\end{abstract}

\section{Introduction}\label{Intro}
One who has a look at verification research and practice may observe that there exist \emph{verification in large (scale)} and \emph{verification in small (scale)}:
verification in large deals (usually) behavioral properties of large-scale complex critical systems like the \emph{Curiosity} Mars mission \cite{Holzmann14},
while verification in small addresses (usually) functional properties of small programs like computing the standard trigonometry functions \cite{Harrison00}.
Verification of behavioral properties of a safety/mission/avalability-critical system doesn't guaranty safety/liveness/fair\-ness of the system
but may detect some bugs that may cause a very expensive and/or fatal system failure
(like launch failure from launch-site ``Vostochny'' November 28, 2017, \cite{Roscosmos}).
At the same time verification in small also is of the high importance:
a \emph{tiny} bug/mistake/error in a \emph{small} but \emph{frequently/massively} used function/program may cause a huge money losses;
it is true in particular for the standard computer functions (available in the standard libraries) \cite{Grohoski17}.
Of course this division of the verification research onto two streams --- in large and in small scale ---
is just a split not a break because all verification research work altogether towards incorporation of the formal verification into
the software development cycle --- at compilation/linking stages maybe \cite{Hoare03}.

Our paper deals with \emph{verification in small}, in particular, it looks like that it is about the same topic as \cite{Harrison00} i.e.
formal verification of two standard computer functions \texttt{cos} and \texttt{sin} that implement well-known  trigonometry mathematical real functions
$\cos,\sin:\mathbb{R}\rightarrow\mathbb{R}$.
But there are serious differences between \cite{Harrison00} and our paper.
Firstly, the cited paper is platform-dependent (Intel\circledR IA-64 architecture), its approach is neither incremental nor combined;
next it provides neither definition of the both $\cos$ and $\sin$ functions, nor specification of their computer partners \texttt{cos} and \texttt{sin};
finally, because of use of HOL-light, all algorithms in the cited paper are functional but not imperative.
In contrast, in our paper we present platform-independent and incremental approach,
based on provided formal definition for mathematical functions, discuss several variants of formal specifications for their computer partners,
use Hoare total correctness assertions \cite{AptDeBoerOlderog09,Gries81} for logical specification of imperative algorithms that implements the computer functions,
and finish with manual (pen-and-paper) verification (using Floyd-Hoare approach \cite{AptDeBoerOlderog09,Gries81}) of the computer functions
for argument value in the rage $[-1,1]$ (in radian measure).
(Thus we postpone computer-aide validation of our proof for the future while the paper \cite{Harrison00} have done computer-aided formal verification.)

Our present paper is a next one in a series of our papers devoted to
the development of a platform-independent incremental combined approach to specification and verification of the standard mathematical functions
\cite{Shilov15,ShilovPromsky16,ShilovAnureevBerdyshevKondratyevPromsky18,ShilovKondratyevAnureevBodinPromsky18}.
Position papers \cite{Shilov15,ShilovPromsky16} have stated our concern regarding a need of
\begin{itemize}
\item better specification and incremental combined platform-independent verification of standard functions,
\item introduction and standardization of a certification process for the standard functions,
\item inclusion of an incremental combined platform-independent verification into this certification.
\end{itemize}

A work-in-progress electronic preprint \cite{ShilovAnureevBerdyshevKondratyevPromsky18}
has presented a human-oriented specification and pen-and-paper verification of a computer square root function
that implements Newton-Raphson method by \emph{non-adaptive} \texttt{for}-loop (with a pre-computed number of iterations)
and uses a look-up table for initial approximations.
The specification in \cite{ShilovAnureevBerdyshevKondratyevPromsky18} has been presented as a total correctness assertion
with use of precise arithmetic and the mathematical square root $\sqrt{\dots}$,
algorithms has been presented by imperative pseudo-code with explicit distinction between precise and machine arithmetic,
manual verification has been done in Floyd-Hoare style and adjustment (matching) of runs of algorithms with precise arithmetics and with machine arithmetics.
It is possible to say that the primary contribution of the paper \cite{ShilovAnureevBerdyshevKondratyevPromsky18}
was an axiomatisation of properties of a machine (fix-point as well as floating-point) arithmetic
that are sufficient to carry out the verification.

A journal (Russian) paper \cite{ShilovKondratyevAnureevBodinPromsky18} is based on an improved axiomatization from \cite{ShilovAnureevBerdyshevKondratyevPromsky18}.
In the cited paper an \emph{adaptive} imperative algorithm implementing the same Newton-Raphson method for a square root function
has been specified by total correctness assertions and verified manually using Floyd-Hoare approach in both fix-point and floating-point arithmetics;
the post-condition of the total correctness assertion states that the final overall error is not greater that $2ulp$
where $ulp$ is \emph{Unit in the Last Place} --- the unit of the last meaningful digit.
The paper \cite{ShilovKondratyevAnureevBodinPromsky18} has reported two steps towards computer-aided validation and verification
of the used adaptive algorithm:
\begin{itemize}
\item an implementation of a fix-point data type according to the axiomatization can be found at
\url{https://bitbucket.org/ainoneko/lib_verify/src/},
\item ACL2 proofs of
\begin{itemize}
\item the consistency of the computer arithmetics axiomatization,
\item the existence of a look-up table with initial approximations for $\sqrt{\dots}$
\end{itemize}
can be found at \url{https://github.com/apple2-66/c-light/tree/master/experiments/square-root}.
\end{itemize}

The paper is organized as follows.
In the next section \ref{WhatAreCoSin} we discuss and recall definitions and properties of the constant $\pi$ and two trigonometric functions
$\cos\equiv\ \lambda x\in\mathbb{R}. \left(\cos x\right)$ and $\sin\equiv\ \lambda x\in\mathbb{R}. \left(\sin x\right)$.
Then in the section \ref{CompPiInReal} we discuss how to compute, specify, and verify approximations of $\pi$ in the real arithmetics;
this section serves as a bridge to the section \ref{CompCoSinInReal} where we
discuss computation, specification, and verification of approximations for values the trigonometric functions $\cos$ and $\sin$ in the real arithmetics.
(Also in future research we will need a verified algorithm to compute approximations of $\pi$.)
In the section \ref{ComputingCoSin} we present some shocking experiments with \emph{direct} implementation on conventional computers
(that use floating-point arithmetic) of algorithms presented and verified (for the real arithmetic) in the previous section  \ref{CompCoSinInReal}:
it turns out that the trigonometric functions $\cos$ and $\sin$ can get values out of the range $[-1,1]$ even for moderate argument values!
Due to these shocking experiments we turn to fix-point arithmetic (and axiomatize it),
design, specify and verify (manually) algorithms to compute approximations of the trigonometric functions $\cos$ and $\sin$
working in fix-point arithmetic with small argument values in the range $[-1,1]$
(i.e. that approximate functions  $\lambda x\in[-1,1]. \left(\cos x\right)$ and $\lambda x\in[-1,1]. \left(\sin x\right)$).
Thus we postpone study of computation, specification, and verification of approximations in floating-point arithmetic
of the functions $\cos$ and $\sin$ for the future;
we discuss in brief this research topic and other future research topics in the last concluding section \ref{concl}.

\section{What are trigonometric functions $\cos$ and $\sin$}\label{WhatAreCoSin}
\begin{figure}[t]
	\centering
	\includegraphics[width=12cm]{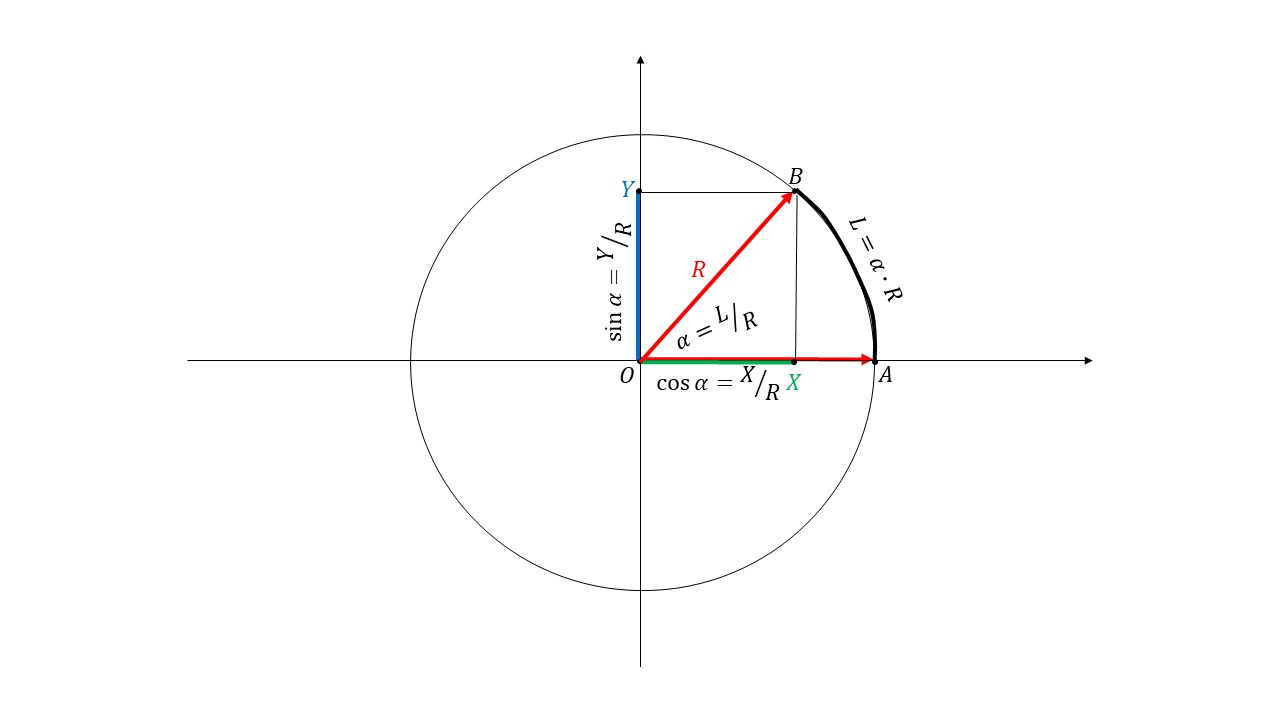}\\
    \caption{\label{CoSinDef} Geometric definition of $\cos\alpha$ and $\sin\alpha$}
\end{figure}
Let us refer Fig.\ref{CoSinDef} to give a geometric definition for the trigonometry functions $\cos,\sin:\mathbb{R}\rightarrow \mathbb{R}$
and some other related definitions.
By definition \cite{Pi}, Archimedes' constant $\pi$ is the ratio of a circle's circumference to its diameter.
Also by definition \cite{RadianInEncyclopedia}, radian measure of the angle between rays defined by radius-vectors $\overrightarrow{OA}$ and $\overrightarrow{OB}$
is $\alpha=\frac{L}{R}$ where $L$ is the length of the arc $\stackrel{\smallfrown}{AB}$.
Finally, by definition \cite{TrigFuncInEncyclopedia}, cosine $\cos\alpha$ and sine $\sin\alpha$
of the angle between two radius-vectors $\overrightarrow{OA}$ and $\overrightarrow{OB}$ with radian measure $\alpha$
is defined as follows: $\cos\alpha = \frac{X}{R}$ and $\sin\alpha = \frac{Y}{R}$;
the following basic equalities immediately follow from definitions and geometric arguments (Pythagorean theorem and triangle equalities):
\begin{itemize}
\item $\left(\cos\alpha\right)^2 +\left(\sin\alpha\right)^2 = 1$ (or shortly $\cos^2\alpha + \sin^2\alpha = 1$);
\item $\cos\alpha = \cos(-\alpha)$ and $\sin(-\alpha) = -\sin\alpha$;
\item $\cos\alpha = \sin\left(\frac{\pi}{2}+\alpha\right)=\sin\left(\frac{\pi}{2}-\alpha\right)$;
\item $\sin\alpha = -\cos\left(\frac{\pi}{2}+\alpha\right)=\cos\left(\frac{\pi}{2}-\alpha\right)$;
\item $\cos\alpha = -\cos(\alpha+\pi)=\cos(\alpha+2\pi)$;
\item $\sin(\alpha) = -\sin(\alpha+\pi)=\sin(\alpha+2\pi)$.
\end{itemize}

More complicated are the following trigonometric addition formulas  \cite{WeissteinAdd}:
\begin{itemize}
\item $\cos(\alpha + \beta) = \cos\alpha \cdot \cos\beta - \sin\alpha \cdot \sin\beta$,
\item $\cos(\alpha - \beta) = \cos\alpha \cdot \cos\beta + \sin\alpha \cdot \sin\beta$,
\item $\sin(\alpha + \beta) = \cos\alpha \cdot \sin\beta + \sin\alpha \cdot \cos\beta$,
\item $\sin(\alpha - \beta) = \cos\alpha \cdot \sin\beta - \sin\alpha \cdot \cos\beta$,
\end{itemize}
where $\alpha$ and $\beta$ are radian measures of two angles.
These formulas imply the following double-angle \cite{WeissteinDbl} and half-angle \cite{WeissteinHlf} formulas
\begin{itemize}
\item $\cos 2\alpha = \cos^2\alpha - \sin^2\alpha$,
\item $\sin 2\alpha = 2\cos\alpha \cdot \sin\alpha$,
\item $\cos \frac{\alpha}{2} = \left(-1\right)^{\lfloor\frac{\alpha+\pi}{2\pi}\rfloor}\sqrt{\frac{1+\cos\alpha}{2}}$,
\item $\sin \frac{\alpha}{2} = \left(-1\right)^{\lfloor\frac{\alpha}{2\pi}\rfloor}\sqrt{\frac{1-\cos\alpha}{2}}$,
\end{itemize}
where $\alpha$ is the radian measures of an angle and $\left(\lambda\ x.\ \lfloor x\rfloor\right):\mathbb{R}\rightarrow\mathbb{Z}$
is the floor function that truncates each real number to the largest integer that isn't greater than the number.

Since $\cos\alpha$ and $\sin\alpha$ are defined for every radian measure $\alpha\in\mathbb{R}$,
one can define two functions $\cos,\sin:\mathbb{R}\rightarrow [-1,1]$ as
$\cos=\left(\lambda \alpha\in\mathbb{R}.\ \cos\alpha\right)$ and $\sin=\left(\lambda \alpha\in\mathbb{R}.\ \sin\alpha\right)$;
remark that in Calculus and Real Analysis the bound variable of the function is usually denoted by $x$, $y$, $z$ instead of $\alpha$ \cite{TrigFuncInEncyclopedia}.
Since our paper is about specification and verification of computations of approximate values of these two trigonometric functions
$\cos$ and $\sin$ we will refer them in this paper as \emph{the} trigonometric functions
(while in more general context the list of trigonometric functions \cite{TrigFuncInEncyclopedia} includes also other functions that can be derived from $\cos$ and $\sin$ ).

It immediately follows from the above basic equalities that these functions satisfy the Pythagorean equality
\begin{equation}\label{Pythagor}
\cos^2x+\sin^2x = 1
\end{equation}
for all $x\in\mathbb{R}$.
It also immediately follows from the same basic equalities, that both functions are periodic with $2\pi$ as the smallest period, i.e.:
\begin{equation}\label{period}
\begin{array}{l}
  \cos\left(x+2\pi\right) = -\cos\left(x+\pi\right) =\cos x, \\
  \sin\left(x+2\pi\right) = \sin\left(x+\pi\right) = \sin x
\end{array}
\end{equation}
for all $x\in\mathbb{R}$.
It follows from the trigonometric addition formulas  that
\begin{equation}\label{addition}
  \begin{array}{l}
   \cos(x + y) = \cos x \cdot \cos y - \sin x \cdot \sin y,  \\
     \sin(x + y) = \cos x \cdot \sin y + \sin x \cdot \cos y
   \end{array}
\end{equation}
for all $x,y\in\mathbb{R}$;
also, it follows from the double-angle formulas that
\begin{equation}\label{double}
\begin{array}{l}
  \cos 2x = \cos^2x - \sin^2x = 2\cos^2x - 1 = 1 - 2\sin^2x, \\
  \sin 2x = 2\cos x \cdot \sin x
     \end{array}
\end{equation}
for all $x\in\mathbb{R}$;
finally, it follows from the double-angle formulas that
\begin{equation}\label{half}
\begin{array}{l}
  \cos \frac{x}{2} = \left(-1\right)^{\lfloor\frac{x+\pi}{2\pi}\rfloor}\sqrt{\frac{1+\cos x}{2}}, \\
  \sin \sin \frac{x}{2} = \left(-1\right)^{\lfloor\frac{x}{2\pi}\rfloor}\sqrt{\frac{1-\cos x}{2}}
     \end{array}
\end{equation}
for all $x\in\mathbb{R}$.

\section{Computing $\pi$ in Reals}\label{CompPiInReal}
Definition for the constant $\pi$  provided in the previous section
aren't convenient to compute their values because of a geometric nature of these definitions.
So we need better ways to compute this constant.

\begin{figure}[t]
  \centering
  \begin{tabular}{|c|c|}
  \hline
  \includegraphics[width=4cm]{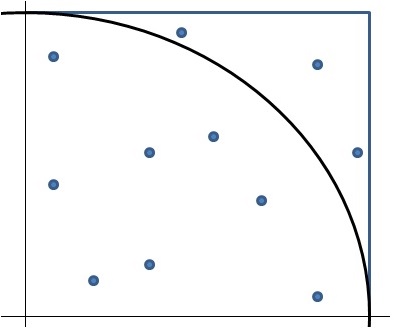}
&
\includegraphics[width=4cm]{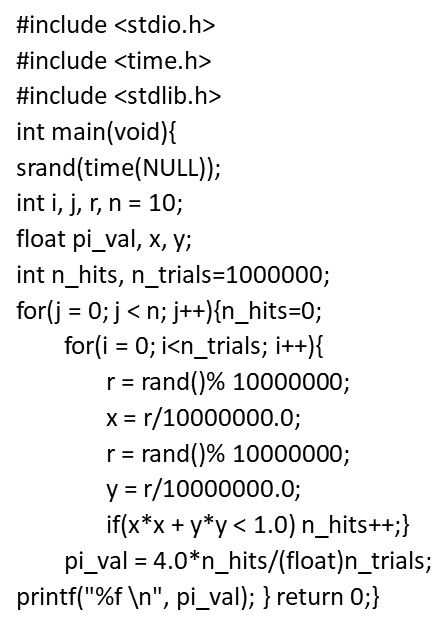}
 \\
  \hline
\end{tabular}
  \caption{Idea (left) and C-code (right) to compute Monte Carlo approximation for $\pi$}\label{IdeaCodePiMonteCalro}
\end{figure}

The following fabulous story is a quotation from the paper \cite{Shilov15}
that was a position paper motivating a need for better specification and verification of the standard mathematical function.
\begin{quote}
\emph{
The mathematical irrational number $\pi$ is the ratio of a  circle's circumference
to its diameter $D$; it is also a well-known mathematical fact that the area of
the circle is
$(\pi \times D^2)/4$, i.e. it is
$\pi/4$ of the area of the square built on the circle's diameter.
This observation leads to the Monte Carlo method for computing an approximation of $\pi$
as follows (Fig. \ref{IdeaCodePiMonteCalro}-left):
draw a segment of a circle in the first quadrant and the square around it,
then randomly place dots in the square;
the ratio of the number of dots inside the circle
to the total number of dots should be approximately equal to $\pi/4$...}

\emph{
The C-program depicted in Fig. \ref{IdeaCodePiMonteCalro}-right  implements the above Monte Carlo method
to compute an approximation for $\pi$.
It prescribes to exercise 10 series of 1,000,000 trials each.
This code was developed by a Computer Science instructor to teach first-year students C-loops
by an example of a very intuitive algorithm.
There were 25 students in the class that used either Code::Blocks 12.11 or Eclipse Kepler IDEs for C/C++ with MinGW environment...}

\emph{
Imagine the embarrassment of the instructor when each of 25 students in the class
got 10 times the value $4.000000$ as an approximation for $\pi$!
}
\end{quote}

Please refer papers \cite{Shilov15,ShilovPromsky16} for detailed discussion what was wrong with the described 25 computational experiments
and for a human ``proof'' that $\pi$ is $4.000000$ indeed.
But we have to rule the idea to compute $\pi$ using Monte Carlo methods because of impossibility to generate random numbers (but pseudo-random only)
that are used in the method.

Instead one can compute approximations for $\pi$ using series --- for example, a Leibnitz' series \cite{Pi}
\begin{equation}\label{PiLeibnitz}
\frac{\pi}{4} = 1 - \frac{1}{3} + \frac{1}{5} - \frac{1}{7} + \dots (-1)^n\frac{1}{2n+1} + \dots = \sum_{n\geq 0} (-1)^n\frac{1}{2n+1}.
\end{equation}
This series representation for $\pi$ results from series representation of the arctangent function
$$\arctan x = \sum_{n\geq 0} (-1)^n\frac{x^{2n+1}}{2n+1}$$
that is valid for all for arguments $x\in[-1,1]$.
The series \ref{PiLeibnitz} converges according to Leibnitz criterion for the alternating series \cite{WeissteinLeibnitz} but slowly:
for every $n\in\mathbb{N}$ its convergence rate is
\begin{equation}\label{PiConvRate}
\left|\frac{\pi}{4} - \left(1 - \frac{1}{3} + \frac{1}{5} - \frac{1}{7} + \dots (-1)^n\frac{1}{2n+1}\right)\right| \leq \frac{1}{2(n+1)+1};
\end{equation}
for example, if we would like to get accuracy $0.001$ for $\pi$ we should compute and summarise (almost) 2,000 terms of the series!

\begin{figure}[t]
\centering
\begin{tabular}{|c|l|}
  \hline
  \multirow{9}{*}{\includegraphics[width=5.5cm]{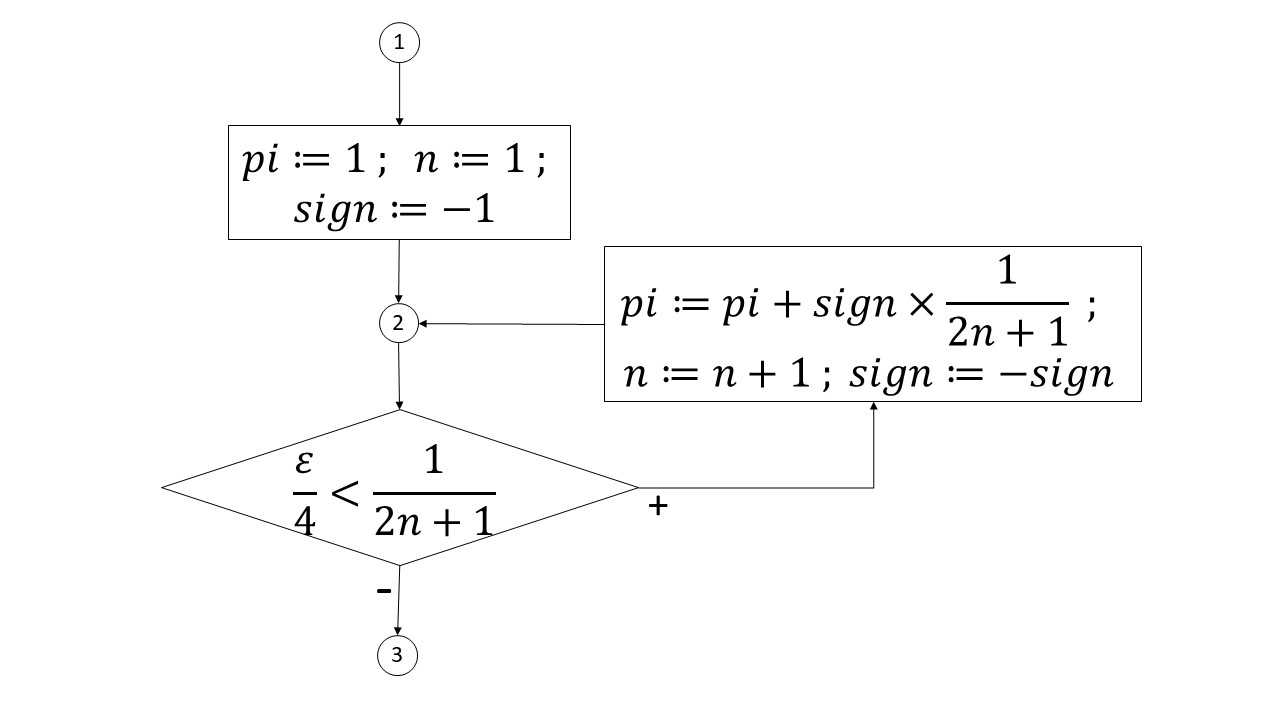}}
  &  \\
    & \\
    & $qp:= 1\ ;\ n:=1\ ;\ sign:=-1\ ;$ \\
    & $while\ \frac{\varepsilon}{4} < \frac{1}{2n+1}\ do$ \\
    & $\hspace*{3em} qp:= qp + sign\times\frac{1}{2n+1}\ ;$ \\
    & $\hspace*{3em} n:=n+1\;\ sign:=-sign\ od$ \\
    & $pi:= 4\times qp$ \\
    & \\
    & \\
  \hline
\end{tabular}
\caption{A flowchart (left) and the pseudo-code (right) of the algorithm $PiCodeInReals$}\label{PiCodeInReals}
\end{figure}

Nevertheless Leibnitz' series gives a way to define $\pi$ and compute its valid approximations using exact real arithmetics:
\begin{itemize}
\item an algorithm $PiCodeInReals$ in Fig. \ref{PiCodeInReals} ``inputs'' accuracy $\varepsilon>0$,
uses variables capable to store (exactly) real numbers  and exercise (exactly) the standard real operations,
and ``outputs'' an approximation $\pi$ with (at least) this accuracy $\varepsilon$;
\item total correctness assertion that specifies the algorithm is
\begin{equation}\label{PiSpecInReals}
\left[\varepsilon>0\right]\ PiCodeInReals\ \left[\left|\pi-pi\right|\leq\varepsilon\right]
\end{equation}
where $\pi$ is defined by (\ref{PiLeibnitz}).
\end{itemize}

It is easy to verify manually the total correctness assertion (\ref{PiSpecInReals}):
\begin{itemize}
\item if $\varepsilon>0$, then the algorithm terminates after $\lceil\frac{2}{\varepsilon} - \frac{3}{2}\rceil$ iteration of the loop $(2+2)$,
where $\left(\lambda\ x.\ \lceil x\rceil\right):\mathbb{R}\rightarrow\mathbb{Z}$
is the ceiling function that rounds up each real number to the least integer that isn't smaller than the number;
\item partial correctness of the algorithm with respect to the same pre-condition and the postcondition can be proved
by Floyd-Hoare method \cite{AptDeBoerOlderog09,Gries81} with
$$precondition\ \&\ sign= (-1)^n\ \&\ \left(qp = \sum_{m=0}^{m=(n-1)} (-1)^m\frac{1}{2m+1}\right)$$
as the invariant of the control point $2$ and using (\ref{PiConvRate}) to prove the exit path $(2-3)$.
\end{itemize}

\section{Computing the trigonometric functions in Reals}\label{CompCoSinInReal}
As for the constant $\pi$, the definitions for the functions $\cos$ and $\sin$ provided in the section \ref{WhatAreCoSin}
aren't convenient to compute their values because of a geometric nature of these definitions.
So we need better ways to compute these functions.

Both functions are smoothness and can be represented by Taylor series \cite{TrigFuncInEncyclopedia}:
\begin{equation}\label{TrigFuncTaylor}
\begin{array}{l}
  \cos x = 1 - \frac{x^2}{2!} + \frac{x^4}{4!} + \dots (-1)^n\frac{x^{2n}}{(2n)!} + \dots = \sum_{n\geq 0}(-1)^n\frac{x^{2n}}{(2n)!}, \\
  \sin x = x - \frac{x^3}{3!} + \frac{x^5}{5!} + \dots (-1)^n\frac{x^{2n+1}}{(2n+1)!} + \dots = \sum_{n\geq 0}(-1)^n\frac{x^{2n+1}}{(2n+1)!}.
\end{array}
\end{equation}
Both series are alternating absolutely converging series for all $x\in\mathbb{R}$,
and --- moreover --- according to Leibnitz criterion for the alternating series \cite{WeissteinLeibnitz},
for every $n\in\mathbb{N}$, if $|x|\leq 2n$ then
\begin{equation}\label{CoSinLeibnitz}
\begin{array}{l}
  \left|\cos x - \left( 1 - \frac{x^2}{2!} + \frac{x^4}{4!} + \dots (-1)^n\frac{x^{2n}}{(2n)!}\right)\right| \leq \left|\frac{x^{2n+2}}{(2n+2)!}\right|, \\
  \\
  \left|\sin x - \left( x - \frac{x^3}{3!} + \frac{x^5}{5!} + \dots (-1)^n\frac{x^{2n+1}}{(2n+1)!}\right)\right| \leq \left|\frac{x^{2n+3}}{(2n+3)!}\right|.
\end{array}
\end{equation}

Very similar to the way to compute $\pi$ approximations, the series (\ref{CoSinLeibnitz} give a way to define functions $\cos$ and $\sin$
and compute their approximate values using exact real arithmetics.
\begin{description}
\item[Cosine:]\mbox{}
\begin{itemize}
\item an algorithm $CosCodeInReals$ in Fig. \ref{CosCodeInReals} ``inputs'' argument value $x$ and accuracy $\varepsilon>0$,
uses variables capable to store (exactly) real numbers  and exercise (exactly) the standard real operations,
and ``outputs'' an approximation for $\cos x$ with (at least) this accuracy $\varepsilon$;
\begin{figure}[t]
\centering
\begin{tabular}{|c|l|}
  \hline
  \multirow{9}{*}{\includegraphics[width=6cm]{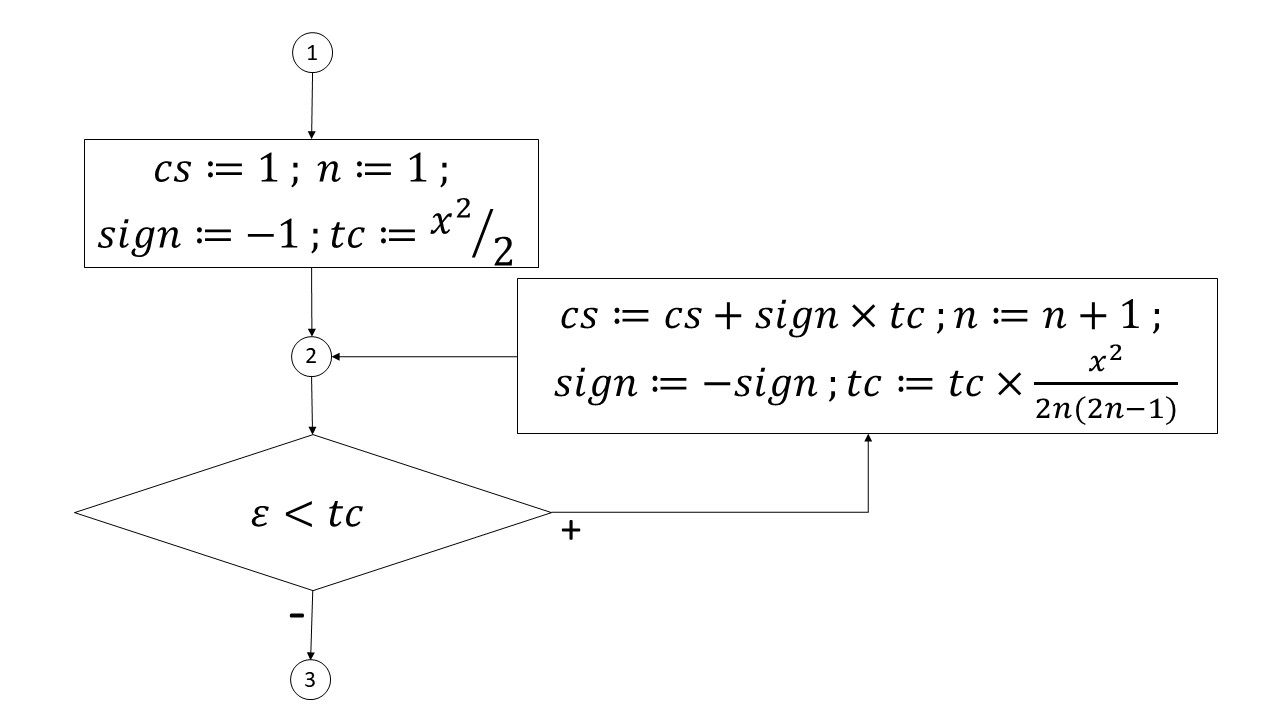}}
  &  \\
  & $cs:= 1\ ;\ n:=1\ ;\ sign:=-1\ ;$ \\
    & $tc:=\frac{x^2}{2}\ ;$ \\
    & $while\ \varepsilon < tc\ do$ \\
    & $\hspace*{3em} cs:= cs + sign\times tc\ ;$ \\
    & $\hspace*{3em} n:=n+1\ ;\ sign:=-sign\ ;$\\
    & $\hspace*{3em} tc:=tc\times\frac{x^2}{2n(2n-1)}\ od$\\
    & \\
    & \\
  \hline
\end{tabular}
\caption{A flowchart (left) and the pseudo-code (right) of the algorithm $CosCodeInReals$}\label{CosCodeInReals}
\end{figure}
\item total correctness assertion that specifies the algorithm is
\begin{equation}\label{CosSpecInReals}
\left[0<\varepsilon<1\right]\ CosCodeInReals\ \left[\left|cs - \cos x\right|\leq\varepsilon\right]
\end{equation}
where $\cos x$ is defined by (\ref{CoSinLeibnitz}).
\end{itemize}
\item[Sine:]\mbox{}
\begin{itemize}
\item an algorithm $SinCodeInReals$ in Fig. \ref{SinCodeInReals} ``inputs'' argument value $x$ and accuracy $\varepsilon>0$,
uses variables capable to store (exactly) real numbers  and exercise (exactly) the standard real operations,
and ``outputs'' an approximation for $\sin x$ with (at least) this accuracy $\varepsilon$;
\begin{figure}[b]
\centering
\begin{tabular}{|c|l|}
  \hline
  \multirow{9}{*}{\includegraphics[width=6cm]{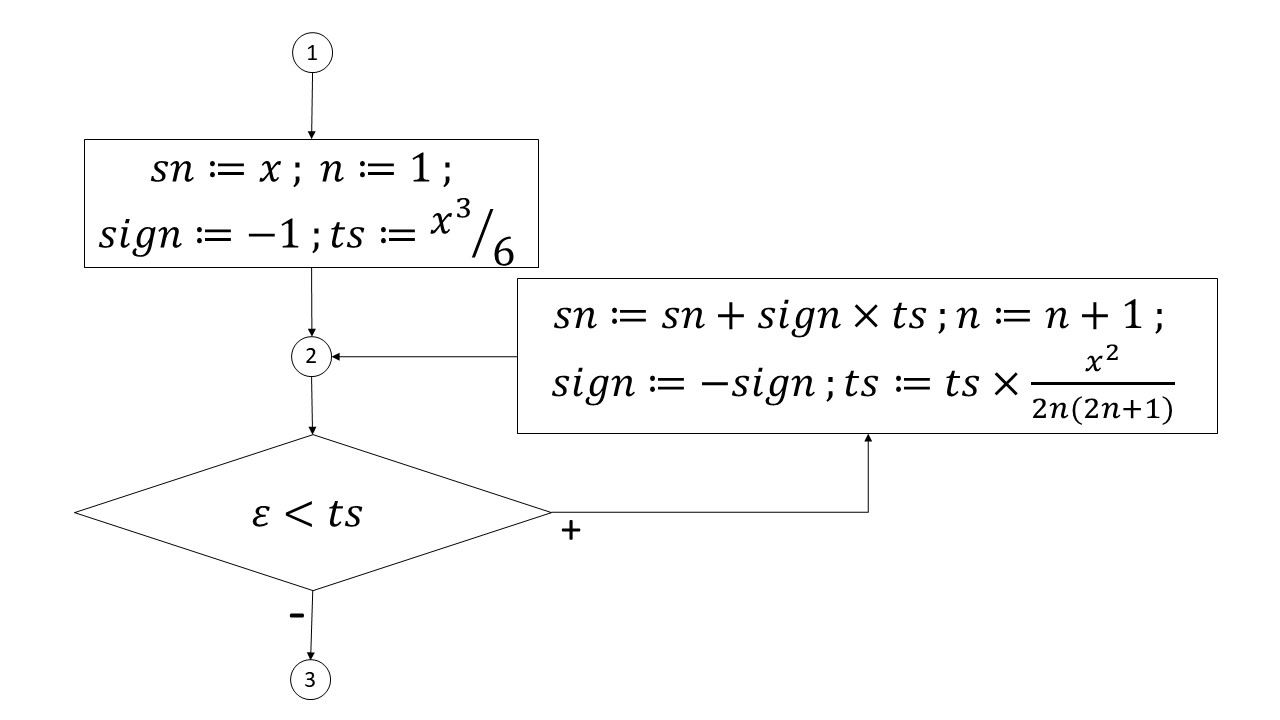}}
  &  \\
  & $sn:= x\ ;\ n:=1\ ;\ sign:=-1\ ;$ \\
    & $ts:=\frac{x^3}{6}\ ;$ \\
    & $while\ \varepsilon < ts\ do$ \\
    & $\hspace*{3em} sn:= sn + sign\times ts\ ;$ \\
    & $\hspace*{3em} n:=n+1\ ;\ sign:=-sign\ ;$\\
    & $\hspace*{3em} ts:=ts\times\frac{x^2}{2n(2n+1)}\ od$\\
    & \\
    & \\
  \hline
\end{tabular}
\caption{A flowchart (left) and the pseudo-code (right) of the algorithm $SinCodeInReals$}\label{SinCodeInReals}
\end{figure}
\item total correctness assertion that specifies the algorithm is
\begin{equation}\label{SinSpecInReals}
\left[0<\varepsilon<1\right]\ SinCodeInReals\ \left[\left|cs - \sin x\right|\leq\varepsilon\right]
\end{equation}
where $\sin x$ is defined by (\ref{CoSinLeibnitz}).
\end{itemize}
\end{description}

Manual verification of the total correctness assertions (\ref{CosSpecInReals}) and \ref{SinSpecInReals}) is very similar to each other
and similar (a little bit more complicated) to the verification of total correctness assertion \ref{PiSpecInReals}) presented at the end of the previous section.
Due to this reason let us discuss below verification of the condition (\ref{CosSpecInReals}) only because of
a similarity of verification of (\ref{CosSpecInReals}) and \ref{SinSpecInReals}).

First let us prove partial correctness of the algorithm $CosCodeInReals$ with respect to the same pre-condition and the postcondition
as in the total correctness assertion (\ref{CosSpecInReals}) using Floyd-Hoare method \cite{AptDeBoerOlderog09,Gries81}.
For it let us adopt the following conjunction
\begin{equation}\label{CosInvInReals}
\left\{\begin{array}{l}
         0<\varepsilon<1 \\
         sign= (-1)^n\\
         tc= \frac{x^{2n}}{(2n)!} \\
         cs = \sum_{m=0}^{m=(n-1)} (-1)^m\frac{x^{2m}}{(2m)!}
       \end{array}
\right.
\end{equation}
as the invariant of the control point $2$.
Pathes $(1..2)$ and $(2+2)$ are easy to prove.
Proof of the exit path $(2-3)$ follows from the convergence rate (\ref{CoSinLeibnitz}) for $\cos$-function,
but we should take care about the applicability condition of the convergence rate: $|x|\leq 2n$;
this condition holds on this path since $1> \varepsilon \geq tc = \frac{x^{2n}}{(2n)!}$
implies that $1> \frac{|x|}{2m}$ or $1> \frac{|x|}{2m-1}$ for some $m\in[1..n]$ and, hence, $1> \frac{|x|}{2n}$.

Next let us prove termination. Let us fix ``input'' (i.e. initial) values  $x$ and $\varepsilon>0$.
Let $m=\min\ n.\left(2n>|x|\right)$, $p=\frac{x^{2m}}{(2m)!}$, and let $k=\lceil\frac{\log_2 p - \log_2\varepsilon}{2}\rceil$.
According to the invariant (\ref{CosInvInReals}), at the $n$-th iteration ($n>0$) of the loop $tc=\frac{x^{2n}}{(2n)!}$;
hence at the $(m+k)$-th iteration of the loop
\begin{center}
$tc=\frac{x^{2m}}{(2m)!}\times \frac{x^{2k}}{\prod_{i=1}^{i=k}\left(\left(2m+(2i-1)\right)(2m+2i)\right)} =$ \hspace*{\fill} \\
$= p\times \prod_{i=1}^{i=k}\frac{x^2}{\left(2m+(2i-1)\right)(2m+2i)} < p\times \prod_{i=1}^{i=k}\left(\frac{1}{2}\right)^2 =$\\
\hspace*{\fill} $= p\times \left(\frac{1}{2}\right)^{2k} \leq p\times\frac{\varepsilon}{p} = \varepsilon$,
\end{center}
i.e. after this iteration the algorithm terminates.

\section{Computing the trigonometric functions on conventional computers}\label{ComputingCoSin}
The previous section ends on a major note: Taylor series definition of the mathematical functions $\cos$ and $\sin$ is easy to
\begin{itemize}
\item implement for approximate computations of values of these functions;
\item use in the specification and verification of these implementations.
\end{itemize}
But there are some objections against being too optimistic:
\begin{itemize}
\item ``easy implementation'' assumes ability of store and manipulate (using the standard arithmetic operations) real numbers;
\item ``easy verification'' is based on a preliminary knowledge from Real Analysis and relay upon human logic reasoning skills.
\end{itemize}

The most common machine approaches to represent some finite subset of real numbers using \emph{fix-point} and/or \emph{floating-point} formats.
Both concepts can be defined as follows.
\begin{description}
  \item[Fixed-point format] has a fixed number (e.g. $4$) for the sign and digits in the integer part (i.e. before the radix/decimal point)
  and a fix number (e.g. $2$) of digits in the fractional part (i.e. after the radix point)
  of some fixed implicit (e.g. binary, decimal, etc.) positional notation;
  for example, in the specified fix-point format $+003.14$ is an approximation of $\pi$.
  \item[Floating-point format] has a fixed number (e.g. $4$) for the sign and digits of the significant/mantissa (usually integer)
  and a fixed number (e.g. $2$) for the sign and digits of the exponent/mag\-nitu\-de (usually integer)
  represented in some fix-point formats (different maybe) and some fixed implicit  base (also usually integer) of the exponent;
  usually the significant doesn't have leading zeros; for example, $+314\times 10^{-2}$
  may be floating-point representation of an approximation of $\pi$ with the significant $+314$, the exponent $-02$, and $10$ as an implicit exponent base.
\end{description}
Fix-point format is quite common in assembly-level programming languages but is rarely used in high-level programming languages;
instead variants of the floating-point formats are popular in high-level programming languages
(e.g. \texttt{float}, \texttt{double}, and \texttt{long double} in C-language \cite{FloatInC}).

Since series (\ref{CoSinLeibnitz}) for $\cos$ and $\sin$  are very similar
as well as the algorithms Fig. \ref{CosCodeInReals} and \ref{SinCodeInReals} that are based on these series,
and different floating-point types in different programming languages have similar problems with representation of real values,
let us consider and discuss below only a series-based implementation of $\cos$ function in C-language with use of \texttt{float} data type.
Then algorithm $CosCodeInReals$ (Fig. \ref{CosCodeInReals}) can be implemented as function \texttt{CosCodeInC} presented in the left part of the Fig. \ref{CosCodeInC};
in this function the variable \texttt{stc} is used for \emph{signed} $tc$ (and assumed to be equal to $-sign\times tc$),
and the variable \texttt{dn} --- for \emph{doubled} $n$ (and assumed to be equal to $2n$).
\begin{figure}[t]
  \centering
  \begin{tabular}{|l|l|}
   \hline
   \texttt{float CosCodeInC(float x, float eps) }            &  \texttt{int TestCosInC(float min,} \\
   \texttt{$\{$  float cs,stc,doubn;}                        &   \hspace*{\fill} \texttt{float max, float step, float eps)}\\
   \texttt{\hspace*{0.5em} cs=stc=dn=1;}                  &  \texttt{$\{$  float x,c;} \\
   \texttt{\hspace*{0.5em} while (fabs(stc) > eps)}     &   \texttt{\hspace*{0.5em} for (x=min; x<=max; x=x+step)} \\
   \texttt{\hspace*{0.5em} $\{$  sigtc=-sigtc*x*x/(dn*(dn+1));} & \texttt{\hspace*{0.5em} $\{$  c=CosCodeInC(x,eps);}  \\
   \texttt{\hspace*{1.5em} cs=cs+stc; dn+=2; $\}$ }        & \texttt{\hspace*{1.5em} printf("\%e  \%e  $\backslash$n",x,c);  $\}$}\\
   \texttt{\hspace*{0.5em} return cs; $\}$} &   \texttt{\hspace*{0.5em} return 0; $\}$} \\
      \hline
 \end{tabular}
\caption{C-function \texttt{CosCodeInC} (left) implements the algorithm $CosCodeInReals$;
 C-function \texttt{TestCosInC} (right) tests the function \texttt{CosCodeInC}}\label{CosCodeInC}
\end{figure}

Table \ref{TestResForCosInC} presents some test data for \texttt{CosCodeInC} with aid of the testing function \texttt{TestCosInC}
(presented in the right part of the Fig. \ref{CosCodeInC})
for argument values in the range $[0,30]$ (radians) with some fixed step ($0.05$) and accuracy ($10^{-6}$).
One can remark  a shocking results in the table \ref{TestResForCosInC}: computed values of \texttt{CosCodeInC} (that are expected to be values of $\cos$)
are $1.144768$ for $ 1.85$ radians and even $13158.8$ for $29$ radians (that can't be true)!
\begin{table}[t]
  \centering
  \begin{tabular}{|l|l|l|l|l|l|}
  \hline
    \texttt{x} & \texttt{CosCodeInC} & \texttt{x} & \texttt{CosCodeInC} & \texttt{x} & \texttt{CosCodeInC} \\
  \hline
    0.000e+000  & 1.000000e+000 & 1.850e+001  & 1.144768e+000 & 2.900e+001  & 1.315880e+004 \\
    5.000e-002  & 9.987502e-001 & 1.900e+001  & 9.913036e-001 & 2.950e+001  & 3.822034e+003 \\
    1.000e-001  & 9.950042e-001 & 1.950e+001  & 6.106047e-001 & 3.000e+001  & -2.368533e+003 \\
  \hline
  \end{tabular}
  \caption{Some tests data for \texttt{CosCodeInC}}\label{TestResForCosInC}
\end{table}

The main reason of this non-tolerable inaccuracy with computing trigonometric functions using floating-point data formats are well-known \cite{DornMcCracken72}:
while $n\leq x$ the values of terms $\frac{x^n}{n!}$ in the series (\ref{CoSinLeibnitz}) are so big that later addition of small values $\frac{x^m}{m!}$ with $x<<m$
``vanish'' because of normalization of floating point values prior to addition.

There are several ways to overcome the problem with computing values of the trigonometric functions $\cos$ and $\sin$
in floating-point arithmetic \cite{DornMcCracken72}:
\begin{itemize}
\item use series (\ref{CoSinLeibnitz}) to compute values of $\cos$ and $\sin$  for \emph{small} argument values;
\item use
\begin{itemize}
\item either \emph{periodicity} (\ref{period}) to reduce \emph{big} argument values to the small ones,
\item or (for example) \emph{Chebychev polynomials}
\begin{itemize}
\item to compute trigonometric functions for positive \emph{integer} argument values
(i.e. $\cos n$ and $\sin n$ for $n\in\mathbb{N}$)
\item and trigonometric addition formulas (\ref{addition}) to compute trigonometric functions for argument values comprising the integer and fractional parts
(the later is in the range $[0,1)$).
\end{itemize}
\end{itemize}
\end{itemize}

Since in any cases there is a need to compute reliable approximations for the values of the trigonometric functions for small argument values
we concentrate in this paper on computation, specification, and (manual) verification in machine arithmetic for  small argument values
--- namely for $x\in[-1,1]$ --- and postpone the study for big argument values for the future.
Also, since all values are small, in this paper we use not a floating-point but a fix-point format to represent the real numbers
in the range of $[-1,1]$.

We would like to conclude this section with a remark that a move from small argument values to big ones isn't trivial:
we strongly agree with Wikipedia \cite{ComputTrigFuncInWiki} that \emph {the computation of trigonometric functions is a complicated subject,
which can today be avoided by most people because of the widespread availability of computers and scientific calculators}
but strongly disagree that modern tools \emph{provide built-in trigonometric functions for any angle} \cite{ComputTrigFuncInWiki}.
Indeed, in the \emph{C reference} \cite{Creference} functions \texttt{cos} \cite{CosInC} and \texttt{sin} \cite{SinInC}
return values are specified as follows.
\begin{description}
\item[\texttt{cos}:] If no errors occur, the cosine of $arg$ ($\cos(arg)$) in the range $[-1,1]$, is returned.
The result may have little or no significance if the magnitude of $arg$ is large. (until C++11)
\item[\texttt{sin}:] If no errors occur, the sine of $arg$ ($\sin(arg)$) in the range $[-1,1]$, is returned.
The result may have little or no significance if the magnitude of $arg$ is large. (until C99)
\end{description}
These specifications are too much vague since they don't explain neither what are $\cos(arg)$ and $\sin(arg)$
(in particular when these values are irrational), no what is \emph{significance}, nor what is \emph{large magnitude}.

\section{The trigonometric functions in Fix-point Arithmetic}\label{CoSinInFixPointArith}
\subsection{Fix-point Arithmetic}\label{FixPointArithm}
First we axiomatized a platform-independent fix-point arithmetic in the electronic preprint \cite{ShilovAnureevBerdyshevKondratyevPromsky18}
and then improved the initial axiomatization in the journal paper \cite{ShilovKondratyevAnureevBodinPromsky18}.
In the present paper we follow the later version, but explicitly admit that there may be several different fix-point data types simultaneously.

A fix-point data-type (with Gaussian rounding)  $\mathbb{D}$ satisfies the following axioms.
\begin{itemize}
	\item The set of values $Val_\mathbb{D}$ is a finite set of rational numbers $\mathbb{Q}$ (and reals $\mathbb{R}$) such that
	\begin{itemize}
		\item it contains the least $\inf_\mathbb{D}<0$ and the largest $\sup_\mathbb{D}>0$ elements,
		\item altogether with
        \begin{itemize}
        \item all rational numbers in $[\inf_\mathbb{D},\sup_\mathbb{D}]$ with a step $\delta_\mathbb{D}>0$,
		\item all integers $Int_\mathbb{D}$ in the range  $[-\inf_\mathbb{D},\sup_\mathbb{D}]$.
        \end{itemize}
	\end{itemize}
	\item Admissible operations include machine addition $\oplus$, subtraction $\ominus$, multiplication $\otimes$, division $\oslash$,
integer rounding up $\lceil\ \rceil$ and down $\lfloor\ \rfloor$.
	\begin{description}
		\item[Machine addition and subtraction.]
		If the exact result of the standard mathematical addition (subtraction) of two fix-point values falls within the interval $[\inf_\mathbb{D},\sup_\mathbb{D}]$,
		then machine addition (subtraction respective\-ly) of these arguments equals to the result of the mathematical operation
        (and notation $+$ and $-$ is used in this case).
		\item[Machine multiplication and division.]
		These operations return values  that are nearest in $Val_\mathbb{D}$ to the exact result of the corresponding standard mathematical operation:
		for any $x,y\in Val_\mathbb{D}$
		\begin{itemize}
			\item if $x\times y\in Val_\mathbb{D}$ then $x\otimes y = x\times y$;
			\item if $x/y\in Val_\mathbb{D}$ then $x\oslash y = x/y$;
			\item if $x\times y\in [\inf_\mathbb{D},\sup_\mathbb{D}]$ then $|x\otimes y - x\times y|\leq\delta_\mathbb{D}/2$;
			\item if $x/y\in [\inf_\mathbb{D},\sup_\mathbb{D}]$ then $|x\oslash y - x/y|\leq\delta_\mathbb{D}/2$;
		\end{itemize}
		\item[Integer rounding up and down] are defined for all values in $Val_\mathbb{D}$.
	\end{description}
	\item Admissible binary relations include all standard equalities and inequalities (within $[\inf_\mathbb{D},\sup_\mathbb{D}]$)
	denoted in the standard way $=$, $\neq$, $\leq$, $\geq$, $<$, $>$.
\end{itemize}

\subsection{Computing \texttt{cos} in Fix-point Arithmetic}\label{ComCosInFixPoint}
\begin{figure}[t]
\centering
\begin{tabular}{|c|c|}
  \hline
 \includegraphics[width=6cm]{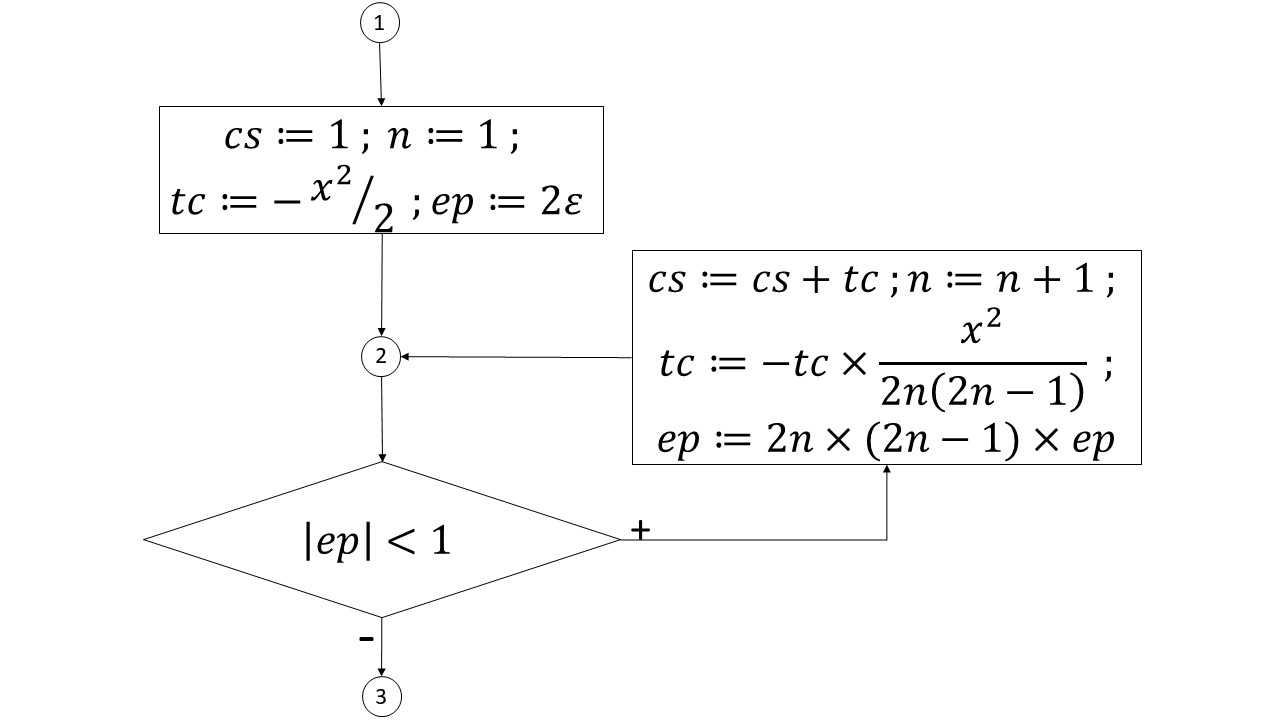}
  &
 \includegraphics[width=6cm]{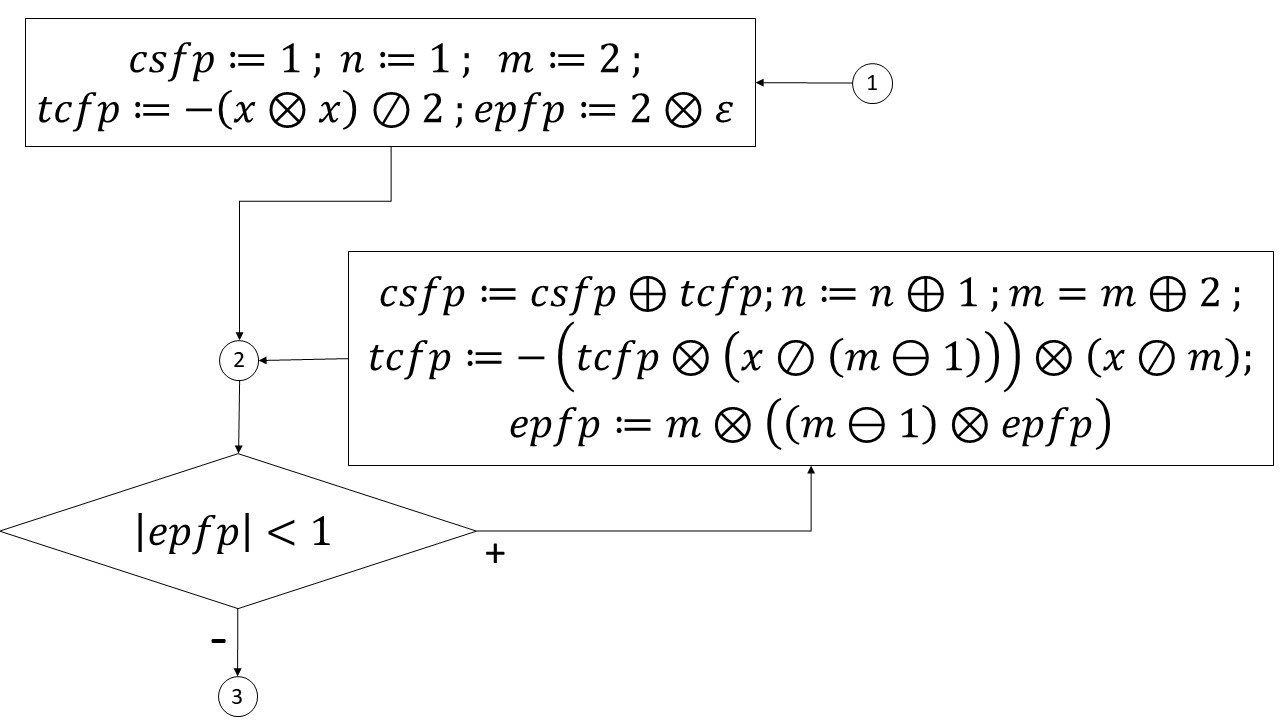}\\
  \hline
\end{tabular}
\caption{Flowcharts of the algorithms $CosCodeInZerOne$ (left) and $CosCodeInFixPoint$ (right)}\label{CosCodeInZerOneFixPoint}
\end{figure}
Since we study the trigonometric functions for argument values in the range $[-1,1]$,
the algorithm $CosCodeInReals$ presented in the Fig. \ref{CosCodeInReals}
can be modified as presented in the left part of the Fig. \ref{CosCodeInZerOneFixPoint};
the corresponding specification (\ref{CosSpecInReals}) should be modified as follows:
\begin{equation}\label{CosSpecInZerOne}
\left[\left(0<\varepsilon<1\right)\ \&\ \left(-1\leq x\leq 1\right)\right]\ CosCodeInZerOne\ \left[\left|cs - \cos x\right|\leq\varepsilon\right]
\end{equation}
where $\cos x$ is defined by (\ref{CoSinLeibnitz}).

Correctness of the specification (\ref{CosSpecInZerOne}) is easy to prove very similar to the proof of the specification (\ref{CosSpecInReals}),
but we need to modify the invariant (\ref{CosInvInReals}) as follows:
\begin{equation}\label{CosInvInZerOne}
\left\{\begin{array}{l}
         0<\varepsilon<1 \\
         -1\leq x\leq 1 \\
         sign= (-1)^n\\
         ep=(-1)^n\times(2n)!\times \varepsilon \\
         tc= \frac{x^{2n}}{(2n)!} \\
         cs = \sum_{m=0}^{m=(n-1)} (-1)^m\frac{x^{2m}}{(2m)!}
       \end{array}
\right.;
\end{equation}
this modified invariant can be used in the proof of the path $(2-3)$ as follows:
$$\left\{\begin{array}{l}
           \left|ep\right|\geq 1 \\
           -1 \leq x \leq 1 \\
           ep=(-1)^n\times(2n)!\times \varepsilon \\
         tc= \frac{x^{2n}}{(2n)!}
         \end{array}\right.\
         \Rightarrow\ \varepsilon = \frac{|ep|}{(2n)!} \geq \frac{1}{(2n)!} \geq \frac{x^{2n}}{(2n)!} = tc;$$
since $tc \geq \left|cs - \cos x\right|$, it implies that $\left|cs - \cos x\right|\leq\varepsilon$.

The above algorithm $CosCodeInZerOne$ presented in the left part of the Fig. \ref{CosCodeInReals}
can be converted into algorithm $CosCodeInFixPoint$ (with fix-point arithmetic) presented in the right part of the Fig. \ref{CosCodeInZerOneFixPoint};
the corresponding specification (\ref{CosSpecInReals}) should be modified as follows:
\begin{equation}\label{CosSpecInFixPoint}
\begin{array}{l c l}
\multicolumn{2}{l}{
\left[
\begin{array}{l}
  \delta_\mathbb{D}< 1\ \& \\
  0<\varepsilon<1\ \&\ \varepsilon\in Val_\mathbb{D}\ \& \\
  \exists N\in Int_\mathbb{D}:\ \left((2N)!\times \varepsilon \geq 1\right)\ \& \\
  -1\leq x\leq 1\ \&\ x\in Val_\mathbb{D}
\end{array}
\right]} & \\
\hspace*{10em} & CosCodeInFixPoint & \hspace*{10em}\\
 &  \multicolumn{2}{r}{\left[
 \begin{array}{l}
   \left|csfp - \cos x\right|\leq\left(\varepsilon + \frac{3n\delta_\mathbb{D}}{2(1-\delta_\mathbb{D})}\right)\ \& \\
   n = \min \left\{N\ :\ \left((2N)!\times \varepsilon \geq 1\right)\right\}
 \end{array}
  \right]}
 \end{array}
\end{equation}
where (as usual in this paper) $\cos x$ is defined  by (\ref{CoSinLeibnitz}).
Please refer Appendix \ref{ProofIdeaSketch} for a proof idea \& sketch for the correctness of this assertion.

\subsection{Computing \texttt{sin} in Fix-point Arithmetic}\label{ComSinInFixPoint}
\begin{figure}[t]
\centering
\begin{tabular}{|c|c|}
  \hline
 \includegraphics[width=6cm]{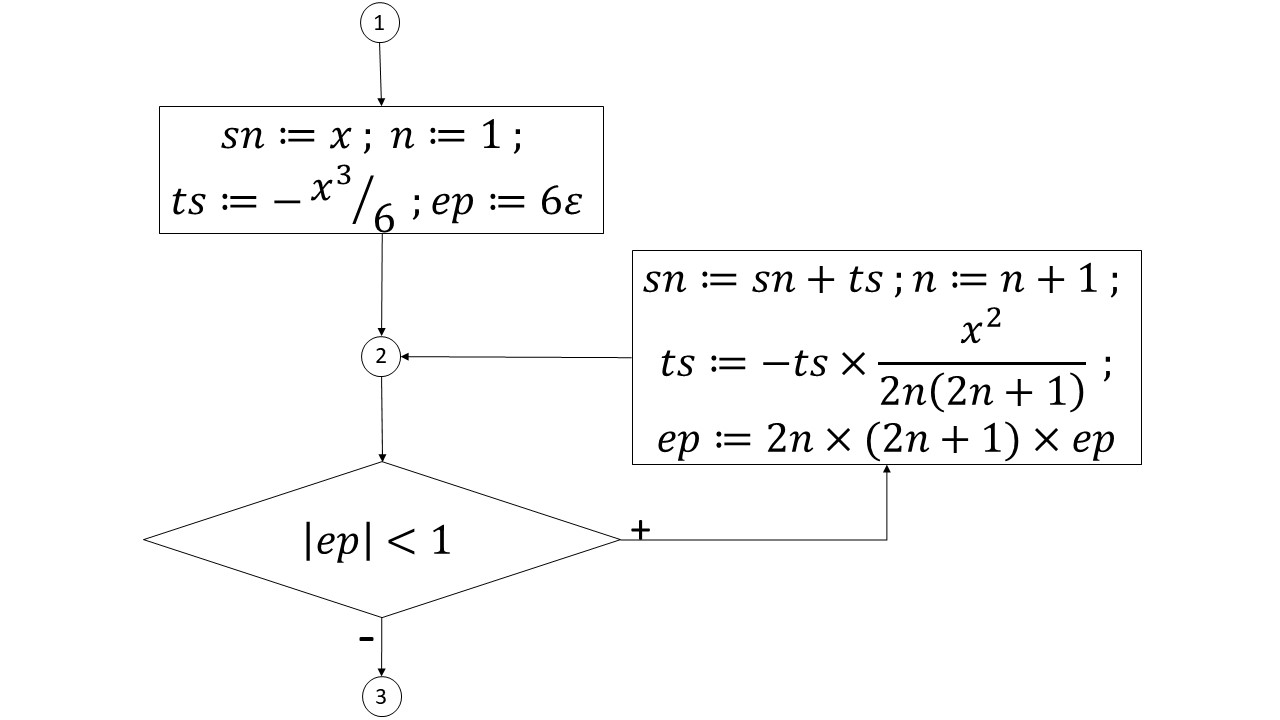}
  &
 \includegraphics[width=6cm]{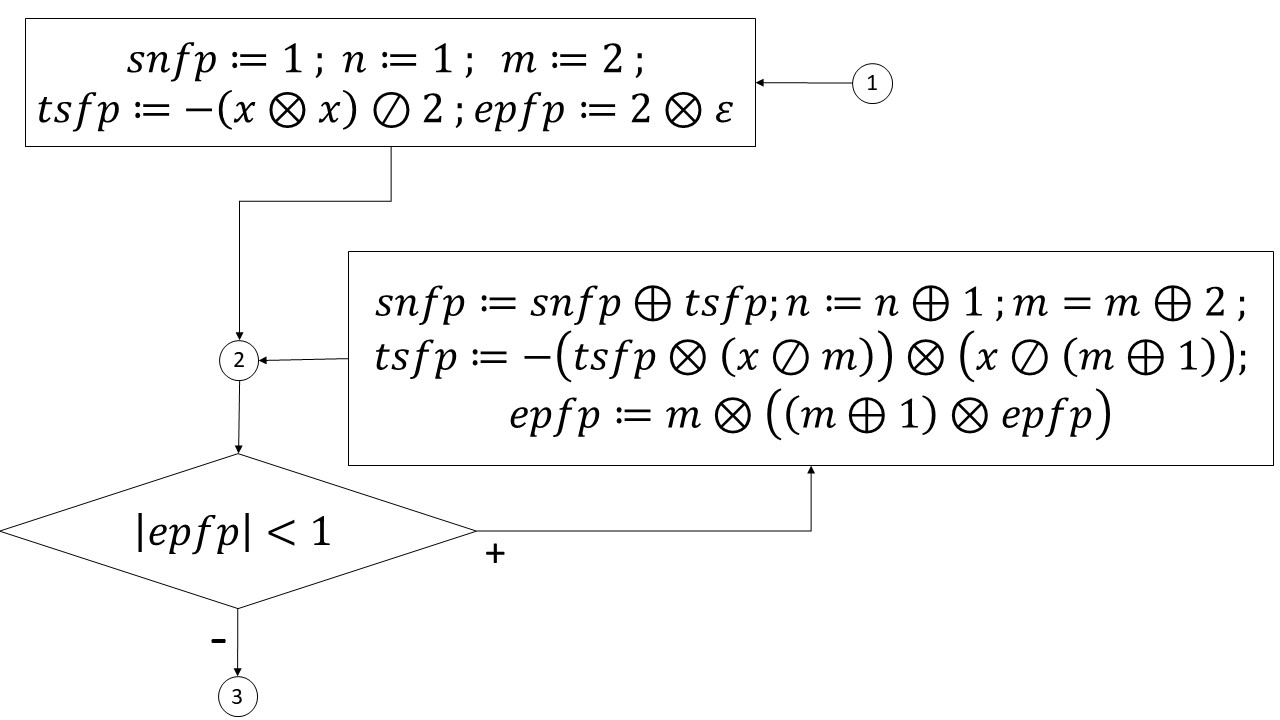}\\
  \hline
\end{tabular}
\caption{Flowcharts of the algorithms $SinCodeInZerOne$ (left) and $SinCodeInFixPoint$ (right)}\label{SinCodeInZerOneFixPoint}
\end{figure}
Like in the above subsection,
the algorithm $SinCodeInReals$ presented in the Fig. \ref{SinCodeInReals}
can be modified for argument values in the range $[-1,1]$ as presented in the left part of the Fig. \ref{SinCodeInZerOneFixPoint};
the corresponding specification (\ref{SinSpecInReals}) should be modified also:
\begin{equation}\label{SinSpecInZerOne}
\left[\left(0<\varepsilon<1\right)\ \&\ \left(-1\leq x\leq 1\right)\right]\ SinCodeInZerOne\ \left[\left|sn - \sin x\right|\leq\varepsilon\right]
\end{equation}
where $\cos x$ is defined by (\ref{CoSinLeibnitz}).
Correctness of the above specification (\ref{SinSpecInZerOne}) can be proved very similar to the proof of the specification (\ref{CosSpecInZerOne}).

Again, similarly to the previous subsection, the above algorithm $SinCodeIn\-Zer\-One$ presented in the left part of the Fig. \ref{SinCodeInReals}
can be converted into algorithm $SinCodeIn\-FixPoint$ (with fix-point arithmetic) presented in the right part of the Fig. \ref{SinCodeInZerOneFixPoint};
the corresponding specification (\ref{SinSpecInReals}) should be modified as follows:
\begin{equation}\label{SinSpecInFixPoint}
\begin{array}{l c l}
\multicolumn{2}{l}{
\left[
\begin{array}{l}
  \delta_\mathbb{D}< 1\ \& \\
  0<\varepsilon<1\ \&\ \varepsilon\in Val_\mathbb{D}\ \& \\
  \exists N\in Int_\mathbb{D}:\ \left((2N+1)!\times \varepsilon \geq 1\right)\ \& \\
  -1\leq x\leq 1\ \&\ x\in Val_\mathbb{D}
\end{array}
\right]} & \\
\hspace*{10em} & SinCodeInFixPoint & \hspace*{10em}\\
 &  \multicolumn{2}{r}{\left[
 \begin{array}{l}
   \left|snfp - \sin x\right|\leq\left(\varepsilon + \frac{3n\delta_\mathbb{D}}{2(1-\delta_\mathbb{D})}\right)\ \& \\
   n = \min \left\{N\ :\ \left((2N+1)!\times \varepsilon \geq 1\right)\right\}
 \end{array}
  \right]}
 \end{array}
\end{equation}
where (as usual in this paper) $\sin x$ is defined  by (\ref{CoSinLeibnitz}).
A proof of this specification should be similar to the proof of the specification  (\ref{CosSpecInFixPoint}).

\section{Conclusion}\label{concl}
In this paper we concentrate on design, specification and (a preliminary manual)
verification of two trigonometric functions $\cos$ and $\sin$ in platform-independent
fix-point arithmetic for small argument values in the range $[-1,1]$ and use Taylor expansions as the definitions of the functions.
Let us enumerate below some problems that need further theoretical and experimental research.

First, we should try to implement our verified algorithms on the virtual computer (for our fix-point arithmetic)
available at \url{https://bitbucket.org/ainoneko/lib_verify/src/} and then test these implementations
against
\begin{itemize}
\item selected algebraic values for these functions
(for example, $\sin \frac{\pi}{6} = \frac{1}{2}$, $\sin \frac{\pi}{4} = \cos \frac{\pi}{4} = \frac{\sqrt{2}}{2}$,
$\cos \frac{\pi}{6} = \frac{\sqrt{3}}{2}$, etc.) in lines with test approach suggested and explained in \cite{Kuliamin07,Kuliamin10};
\item automatically generated test data computed using Taylor expansions in any language that supports unbounded integer arithmetic;
for example, a Lisp-function in Fig. \ref{LispFunc} computes approximations for $\cos$ in unbounded rational arithmetics with accuracy $10^{-8}$.
\end{itemize}
\begin{figure}[t]
\begin{verbatim}
(defun my-cos (x &optional (eps 1E-8))
  (let ((a 1) (s 1) (k 0))
    (loop
      (when (<= (abs a) eps) (return s))
      (setq a (- (/ (* a x x) (+ k 1) (+ k 2)))
                s (+ s a)
                k (+ k 2)))))
\end{verbatim}
  \caption{Lisp-function to compute approximations for $\cos$ in unbounded rational arithmetic with accuracy $10^{-8}$}\label{LispFunc}
\end{figure}

Next we should complete the section \ref{CoSinInFixPointArith} by a pen-and-paper proof of the specification (\ref{CosSpecInFixPoint})
(instead of the sketch presented in the appendix \ref{ProofIdeaSketch}) and by a proof of the specification (\ref{SinSpecInFixPoint}).
Then we should validate/implement both proofs using some proof-assistance since
manual proofs accompanied by computer-aided proofs is the core idea of the combined approach to verification.
Currently in our studies of the square-root function \cite{ShilovAnureevBerdyshevKondratyevPromsky18,ShilovKondratyevAnureevBodinPromsky18}) 
we  are using  proof-assistance ACL2 for proof-validation/implementation, but may change our choice later.

Finally we should move from computation, specification and verification of approximations of the trigonometric functions for small argument values in fix-point arithmetic
to relatively big argument values in floating-point arithmetics.
As we have mentioned in the section \ref{ComputingCoSin}, computing of the values of the trigonometric functions for big argument values
may be reduced to small argument values  
either using periodicity (\ref{period}), or (for example) Chebychev polynomials, the trigonometric addition \ref{addition},
the double-angle \ref{double}, and the half-angle \ref{half} formulas.
(Remark that in the first case we need to compute approximate values of the constant $\pi$ with high precision.)

We would like to finish the paper with a remark that the test-based approach from \cite{Kuliamin07,Kuliamin10}
may be used for  argument range larger than $[-1,1]$; 
automated testing against valid approximations computed using unbounded rational arithmetic also may help;
for example table \ref{MyCos50} presents rational approximation of $\cos 50$ computed as \texttt{(my-cos 50)} using unbounded rational arithmetic
(this rational value is ``equal'' to a float-point value $0.9649660286$).
\begin{table}[t]
  \centering
  \begin{tabular}{|l|l|}
    \hline
    nominator & denominator \\
    \hline
    24370613165454113267560338608221954 & 25255410493873184332225648114958816 \\
    98255428138520309455467035800407203 & 94660898821193613023561185556763590 \\
    92481216493267961919792183534114282 & 78896631844387898015300688850221053 \\
    43256901695353743984506265611950655 & 37104695728469968259460206109490815 \\
    23779221083103374016633819981723287 & 73617550435820266050926650594970281 \\
    8060581913569126766599 &  3572299506856327202849 \\
    \hline
  \end{tabular}
  \caption{Result of evaluation of \texttt{(my-cos 50)}}\label{MyCos50}
\end{table}

\appendix

\section{Proof Sketch for the Correctness Assertion \ref{CosSpecInFixPoint}}\label{ProofIdeaSketch}
Let us remark that a conjunct
$$\exists N\in Int_\mathbb{D}:\ \left((2N)!\times \varepsilon \geq 1\right)$$
in the precondition of (\ref{CosSpecInFixPoint}) implies that
\begin{itemize}
\item the algorithm $CosCodeInFixPoint$ terminates,
the final value of the variable $n$ is $\min \left\{N\ :\ \left((2N)!\times \varepsilon \geq 1\right)\right\}$ indeed,
it is equal to the number of iterations of the algorithm $CosCodeInFixPoint$ and to number of iterations of the algorithm $CosCodeInZerOne$;
\item in all assignments in $CosCodeInFixPoint$ but the following two
\begin{equation}\label{CritAssInCosCode}
\begin{array}{c}
  tcfp:=-(x\otimes x)\oslash 2 \\
  tcfp:=-\left(tcfp\otimes \left(x\oslash\left(m\ominus 1\right)\right)\right)\otimes \left(x\oslash m\right)
\end{array}
\end{equation}
all the floating-point operations are equal to the corresponding \emph{exact} real arithmetic operations
(e.g. $epfp:= m\otimes\left(\left(m\ominus 1\right)\otimes epfp\right)$ can be replaced by $epfp:=m\times (m-1)\times epfp$);
\item more over, in the second of these assignments (\ref{CritAssInCosCode})
the floating-point subtraction is equal to the  \emph{exact} real arithmetic subtraction, i.e. this assignment can be replaced by
$tcfp:=-\left(tcfp\otimes \left(x\oslash\left(m-1\right)\right)\right)\otimes \left(x\oslash m\right)$.
\end{itemize}

Due to the above observations (remarks) we can exercise both algorithms $CosCodeInZerOne$ and $CosCodeInFixPoint$
synchronously in iterations.
For all integer $k\in[1..\min \left\{N\ :\ \left((2N)!\times \varepsilon \geq 1\right)\right\}]$ let
$tcfp_k$ and $scfp_k$ be instant values of the corresponding variables before $k$-th check of the (equal) conditions
$|ep|<1$ and $|epfp|<1$ in the algorithms  $CosCodeInZerOne$ and $CosCodeIn\-FixPoint$ respectively;
let also $\Delta_k = tcfp_k - tc_k$.
In particular,
\begin{equation}\label{Delta1Theta1}
\Delta_1 = tcfp_1 - tc_1 = -(x\otimes x)\oslash 2 + \frac{x^2}{2};
\end{equation}
the absolute value of $\Delta_1$ may be evaluated a follows:
\begin{equation}\label{AbsDelta1}
\begin{array}{c}
|\Delta_1| =  \left|-(x\otimes x)\oslash 2 + \frac{x^2}{2}\right| =
\left|\left(-\frac{x\otimes x}{2} + \frac{\delta^\prime}{2}\right) + \frac{x^2}{2}\right| = \hspace*{8em}\\
\hspace*{\fill} = \left|\left(-\frac{x^2 + \frac{\delta^{\prime\prime}}{2}}{2} + \frac{\delta^\prime}{2}\right) + \frac{x^2}{2}\right|
= \left|\frac{\delta^\prime}{2} - \frac{\delta^{\prime\prime}}{4}\right| \leq \left|\frac{\delta^\prime}{2}\right| + \left|\frac{\delta^{\prime\prime}}{4}\right|
= \frac{3}{4}\delta_\mathbb{D},
\end{array}
\end{equation}
where $|\delta^\prime|,|\delta^{\prime\prime}|\leq\delta_\mathbb{D}$.

Let $k$ be any integer in the range $[1..\min \left\{N\ :\ \left((2N)!\times \varepsilon \geq 1\right)\right\})$.
As we know (\ref{CosInvInReals}), $tc_k=(-1)^k\times\frac{x^{2k}}{(2k)!}$ and $tc_{k+1}=(-1)^{k+1}\times\frac{x^{2(k+1)}}{(2(k+1))!}=-tc_k\times\frac{x^2}{(2k+1)(2k+2)}$.
For technical convenience let us introduce auxiliary values
\begin{equation}\label{AuxilKP}
\begin{array}{c}
  tc_{k+}=tc_k\times\frac{x}{2k+1}, \\
  tcfp_{k+}=tcfp_k\otimes \left(x\oslash\left(2k+1\right)\right), \\
  \Delta_{k+} = tcfp_{k+} -tc_{k+};
\end{array}
\end{equation}
then
\begin{equation}\label{AuxilKPP}
\begin{array}{c}
  tc_{k+1}=-tc_{k+}\times\frac{x}{2k+2}, \\
  tcfp_{k+1}=-tcfp_{k+1}\otimes \left(x\oslash\left(2k+2\right)\right), \\
  \Delta_{k+1} = tcfp_{k+1} -tc_{k+1}.
\end{array}
\end{equation}
We have:
\begin{center}
$tcfp_{k+}=tcfp_k\otimes \left(x\oslash\left(2k+1\right)\right) =
\left(tc_k + \Delta_k\right)\otimes\left(\frac{x}{2k+1}+\frac{\delta^\prime}{2}\right) = $ \hspace*{\fill} \\
$= \left(tc_k + \Delta_k\right)\times \left(\frac{x}{2k+1}+\frac{\delta^\prime}{2}\right) + \frac{\delta^{\prime\prime}}{2}=$ \\
\hspace*{\fill} $= tc_{k+} + \frac{x^{2k}}{(2k)!}\times\frac{\delta^\prime}{2} + \Delta_k\times \left(\frac{x}{2k+1}+\frac{\delta^\prime}{2}\right)+
\frac{\delta^{\prime\prime}}{2}$
\end{center}
where $|\delta^\prime|,|\delta^{\prime\prime}|\leq\delta_\mathbb{D}$;
since $|x|\leq 1$ we can evaluate $|\Delta_{k+}|$ in terms of $|\Delta_k|$ and $\delta_\mathbb{D}$ as follows:
\begin{equation}\label{AbsDeltaKP}
|\Delta_{k+}| =\left|\frac{x^{2k}}{(2k)!}\times\frac{\delta^\prime}{2} + \Delta_k\times \left(\frac{x}{2k+1}+\frac{\delta^\prime}{2}\right)+
\frac{\delta^{\prime\prime}}{2}\right| \leq \frac{1+\delta_\mathbb{D}}{2}|\Delta_{k}| + \frac{3}{4}\delta_\mathbb{D}.
\end{equation}
Due to similarity between (\ref{AuxilKP}) and (\ref{AuxilKPP}) we can evaluate $|\Delta_{k+1}|$ in terms of $|\Delta_{k+}|$ and $\delta_\mathbb{D}$ as follows:
\begin{equation}\label{AbsDeltaKPP}
|\Delta_{k+1}| \leq \frac{1+\delta_\mathbb{D}}{2}|\Delta_{k+}| + \frac{3}{4}\delta_\mathbb{D}.
\end{equation}
Combining (\ref{AuxilKP}) and (\ref{AuxilKPP}) we get
\begin{equation}\label{AbsDeltaKPOne}
|\Delta_{k+1}| \leq \left(\frac{1+\delta_\mathbb{D}}{2}\right)^2\times|\Delta_{k}| + \left(\frac{1+\delta_\mathbb{D}}{2}+1\right)\times\frac{3}{4}\delta_\mathbb{D}.
\end{equation}
Inequalities (\ref{AbsDelta1}) and (\ref{AbsDeltaKPOne}) together imply that
\begin{equation}\label{AbsDeltaOneKPOne}
\begin{array}{c}
|\Delta_{k+1}| \leq \left(\frac{1+\delta_\mathbb{D}}{2}\right)^{2k}\times|\Delta_{1}| +
\left(\sum_{i=0}^{i=2k-1}\left(\frac{1+\delta_\mathbb{D}}{2}\right)^i\right)\times\frac{3}{4}\delta_\mathbb{D} \leq \hspace*{\fill} \\
\hspace*{6em} \leq \left(\sum_{i=0}^{i=2k}\left(\frac{1+\delta_\mathbb{D}}{2}\right)^i\right)\times\frac{3}{4}\delta_\mathbb{D} =
\frac{\left(\frac{1+\delta_\mathbb{D}}{2} \right)^{2k+1}-1}{\frac{1+\delta_\mathbb{D}}{2}-1}\times\frac{3}{4}\delta_\mathbb{D} = \\
\hspace*{\fill} = \frac{1-\left(\frac{1+\delta_\mathbb{D}}{2} \right)^{2k+1}}{1-\delta_\mathbb{D}}\times\frac{3}{2}\delta_\mathbb{D}
\leq \frac{3}{2} \frac{\delta_\mathbb{D}}{1-\delta_\mathbb{D}}.
\end{array}
\end{equation}

Using (\ref{AbsDelta1}) and (\ref{AbsDeltaOneKPOne}) we can prove the first property from the postcondition of the specification (\ref{CosSpecInFixPoint}):
\begin{center}
$\left|csfp - \cos x\right| = \left|\sum_{k=1}^{k=n-1}tcfp_k - \left(\sum_{k=1}^{k=n-1}tc_k - \sum_{k=1}^{k=n-1}tc_k\right) - \cos x\right| \leq $ \hspace*{\fill} \\
$\leq \left|\sum_{k=1}^{k=n-1}tcfp_k - \sum_{k=1}^{k=n-1}tc_k\right| + \left|\sum_{k=1}^{k=n-1}tc_k - \cos x\right| \leq $
\hspace*{\fill} $\leq \sum_{k=1}^{k=n-1}|tcfp_k tc_k| + \varepsilon = |\Delta_1| + \leq \sum_{k=2}^{k=n-1}|tcfp_k tc_k| + \varepsilon \leq $ \\
\hspace*{\fill} $\frac{3}{4}\delta_\mathbb{D} + (n-2)\times\left(\frac{3}{2} \frac{\delta_\mathbb{D}}{1-\delta_\mathbb{D}}\right) \leq
\frac{3n\delta_\mathbb{D}}{2(1-\delta_\mathbb{D})}$.
\end{center}


\begin{thebibliography}{9}

\bibitem{AptDeBoerOlderog09}
Apt K.R., de Boer F.S., Olderog E.-R. Verification of Sequential and Concurrent Programs. Springer-Verlag, 2009.

\bibitem{Creference}
C reference. \url{https://en.cppreference.com/w/c}.
(Visited December 27, 2018.)

\bibitem{CosInC}
C reference: cos, cosf, cosl.
\url{https://en.cppreference.com/w/c/numeric/math/cos}.
(Visited December 27, 2018.)

\bibitem{FloatInC}
C reference: Fundamental types.
\url{https://en.cppreference.com/w/cpp/language/types}.
(Visited December 27, 2018.)

\bibitem{SinInC}
C reference: sin, sinf, sinl.
\url{https://en.cppreference.com/w/c/numeric/math/sin}.
(Visited December 27, 2018.)

\bibitem{DornMcCracken72}
Dorn W.S., McCracken D.D.  Numerical Methods with Fortran IV Case Studies. John Wiley \& Sons, 1972.

\bibitem{Gries81}
Gries D. The Science of Programming. Springer-Verlag, 1981.

\bibitem{Grohoski17}
Grohoski G. Verifying Oracle's SPARC Processors with ACL2. Slides of the Invited talk for 14th International Workshop on the
ACL2 Theorem Prover and Its Applications.
\url{http://www.cs.utexas.edu/users/moore/acl2/workshop-2017/slides-accepted/grohoski-ACL2_talk.pdf}.
(Visited December 27, 2018.)

\bibitem{Harrison00}
Harrison J. Formal Verification of Floating Point Trigonometric Functions.
Lecture Notes in Computer Science. 2000. Vol. 1954. P.217--233.

\bibitem{Hoare03}
Hoare C.A.R. The Verifying Compiler: A Grand Challenge for Computing Research.
Lecture Notes in Computer Science. 2003. Vol. 2890. P. 1--12.

\bibitem{Holzmann14}
Holzmann G.J. Mars Code. Commun. ACM, 2014, Vol. 57(2), p. 64--73.

\bibitem{Kuliamin07}
Kuliamin V. Standardization and Testing of Mathematical Functions.
Programming and Computer Software. 2007. Vol. 33(3). P. 154--173.

\bibitem{Kuliamin10}
Kuliamin V.V.
Standardization and Testing of Mathematical Functions in floating point numbers.
Lecture Notes in Computer Science. 2010. Vol. 5947. P. 257--268.

\bibitem{Pi}
Pi(number). Encyclopedia of Mathematics.
\url{http://www.encyclopediaofmath.org/index.php?title=Pi(number)&oldid=43586}.
(Visited December 27, 2018.)

\bibitem{RadianInEncyclopedia}
Radian. Encyclopedia of Mathematics. \url{http://www.encyclopediaofmath.org/index.php?title=Radian&oldid=31576}.
(Visited December 27, 2018.)

\bibitem{Roscosmos}
``Roskosmos'' nazval prichinu neudachnogo zapuska s kosmodroma Vostochnyy (Roskosmos named the cause of the launch failure from launch-site ``Vostochny'').
\url{https://www.rbc.ru/politics/12/12/2017/5a2ebcd59a79479d29667115}.
(In Russian. Visited December 27, 2018.)

\bibitem{Shilov15}
Shilov N.V. On the need to specify and verify standard functions.
The Bulletin of the Novosibirsk Computing Center (Series: Computer Science), 2015, Vol. 38, p. 105--119.

\bibitem{ShilovPromsky16}
Shilov N.V., Promsky A.V. On specification andd verification of standard mathematical functions.
Humanities and Science University Journal, 2016, Vol. 19, p. 57--68.

\bibitem{ShilovAnureevBerdyshevKondratyevPromsky18}
Shilov N.V. Anureev I.S. Berdyshev M., Kondratyev D., Promsky A.V.
Towards platform-independent verification of the standard mathematical functions: the square root function.
\url{https://arxiv.org/abs/1801.00969} [arXiv:abs/1801.00969].
(Visited December 27, 2018.)

\bibitem{ShilovKondratyevAnureevBodinPromsky18}
Shilov N. V., Kondratyev D. A., Anureev I. S., Bodin E. V, Promsky A. V.
Platform-independent Specification and Verification of the Standard Mathematical Square Root Function.
Modeling and Analysis of Information Systems, 2018, Vol. 25(6), p. 637--666.

\bibitem{TrigFuncInEncyclopedia}
Trigonometric functions. V.I. Bityutskov (originator), Encyclopedia of Mathematics.
\url{http://www.encyclopediaofmath.org/index.php?title=Trigonometric_functions&oldid=14919}.
(Visited December 27, 2018.)

\bibitem{ComputTrigFuncInWiki}
Trigonometric functions: Computation. From Wikipedia, the free encyclopedia.
\url{https://en.wikipedia.org/wiki/Trigonometric_functions#Computation}.
(Visited December 27, 2018.)

\bibitem{WeissteinAdd}
Weisstein E. W.
Trigonometric Addition Formulas. From MathWorld--A Wolfram Web Resource. \url{http://mathworld.wolfram.com/Double-AngleFormulas.html}.

\bibitem{WeissteinDbl}
Weisstein E. W.
Double-Angle Formulas. From MathWorld--A Wolfram Web Resource. \url{http://mathworld.wolfram.com/TrigonometricAdditionFormulas.html}.

\bibitem{WeissteinHlf}
Weisstein E. W.
Half-Angle Formulas. From MathWorld--A Wolfram Web Resource. \url{http://mathworld.wolfram.com/Half-AngleFormulas.html}.

\bibitem{WeissteinLeibnitz}
Weisstein E. W.
Leibniz Criterion. From MathWorld--A Wolfram Web Resource. \url{http://mathworld.wolfram.com/LeibnizCriterion.html}.

 \end{thebibliography}
\end{document}